\documentclass{ws-ijmpa}

\begin{document}

\markboth{Sini R, Nijo Varghese and V C Kuriakose } {Absorption
cross section of RN and SdS extremal black hole}

%
\catchline{}{}{}{}{}
%

\title{Absorption cross section of RN and SdS extremal black hole }

\author{SINI R}

\address{Department of Physics, Cochin University of Science and Technology,
Kochi 682022, India.
\\
 sini@cusat.ac.in}
\author{NIJO VARGHESE}

\address{Department of Physics, Cochin University of Science and Technology,
Kochi 682022, India.
\\
 nijovarghese@cusat.ac.in}

\author{V C KURIAKOSE}

\address{Department of Physics, Cochin University of Science and Technology,
Kochi 682022, India.
\\
 vck@cusat.ac.in}

\maketitle

\begin{history}
\received{Day Month Year}
\revised{Day Month Year}
\end{history}

\begin{abstract}
The nature of scalar wave functions near the horizon of Reissner
Nordstrom (RN) extremal and Schwarzschild-de Sitter (SdS) extremal
black holes are found using WKB approximation and the effect of
reflection of waves from the horizon. The absorption cross section
$\sigma _{abs}$ when RN extremal and SdS extremal black holes placed
in a Klein-Gordon field is calculated. 
\end{abstract}

\ccode{PACS numbers: 04.70.Dy, 04.70.-s }
\section{Introduction}
 Black holes are one of the most enigmatic constructs in the present day
physics. On the one hand, they are the most simple predictions of
the general relativity and on the other hand, they are the
paradigmatic objects to test possible quantum theories of gravity.
To obtain a deeper theoretical understanding of these extreme
objects, many authors have considered the scattering of coherent
waves \cite{ss}. If the wave length of the incident wave is
comparable  with the size of the event horizon, then the wave will
be diffracted by the black hole. Diffraction effects are responsible
for many interesting phenomena in nature, so wave scattering from
black holes is an interesting field of its own right, even if
observations may not realizable in practice.

    A considerable effort has taken place in studying the waves
scattered off by black holes. Both numerical and analytical methods
in solving the various wave equations in black hole scattering have
been developed \cite{ss,ii}. Interest in the absorption of quantum
waves by black hole was reignited in the 1970s, following Hawking's
discovery that black holes can emit, as well as scatter and absorb,
radiation \cite{abaa}. Hawking showed that the evaporation rate is
proportional to the total absorption cross section. Unruh \cite{pp}
found the absorption cross section for massive scalar and Dirac
particles scattered off by small non rotating black holes. In a
series of papers Sanchez  \cite{abab,abac,abad} considered the
scattering and absorption of massless scalar particles by an
uncharged, spherically symmetric black hole.

 Another quantum effect of interest is that event horizons need not be fully absorptive
type but can reflect waves falling on it. It is also proposed that
event horizons has a finite energy width. 't Hooft \cite{viv}
explained the horizon of the black hole as a brick wall so that the
outer horizon $r_+$ spreads into a range of
$\left(r_+-\Delta,r_++\Delta\right)$. Quantum horizon concepts were
introduced by Mu-Lin Yan and Hua Bai \cite{22}. The relevant
equation governing a scattering process in a black hole space time
is analogous to Schrodinger type equations governing scattering
phenomena in quantum mechanics. Hence the standard techniques used
to study quantum scattering can be used to study scattering problems
in black hole space time.

 The work of Wang et al.\cite{bw98} shown that a non extremal RN black hole cannot turn into an extremal one by assimilating an in falling charged particle and shell. So in the present work we are interested in studying the scattering of scalar waves in the RN extremal and SdS extremal space-time. Earlier several authors
have studied scattering of scalar and Fermi fields under different
black hole space time and calculated absorption cross sections. In
all these calculations the black hole is assumed to be capable of
absorbing the radiation falling on it, but here we consider that
both absorption and reflection could take place at the horizon of
black holes \cite{sbn,bb}. The absorption cross section, of scalar
waves in Schwarzschild-de Sitter space time, of charged scalar wave
in RN space time and of dirac wave in Schwarzschild space time were
found earlier \cite{sr,sr2,sr3}. In the present work we use WKB
approximation to get the solution of radial wave equation in the
vicinity of the horizon of black holes. In section 2 we explain how
to obtain the absorption cross section of RN extremal black hole in
low energy limit. Section 3 contains a calculation of absorption
cross section scalar wave scattered off by SdS extremal black hole.
Here we take into consider both reflection and absorption properties
of the black hole horizon. Section IV concludes the paper.
\section{RN black hole - Extremal case}
The RN black hole's (event and inner) horizons in terms of the black
hole parameters are given by, $r_{\pm}=M\pm\sqrt{M^{2}-Q^{2}}$,
where $M$ and $Q$ are respectively mass and charge of black hole. In
extreme case, these two horizons coincide. i.e, when, $M=Q$,
$r_{\pm}=M=r_{0}$ . The metric then is given by,
\begin{equation}
\label{a11}ds^2=\left( 1-\frac{r_o}r\right) ^2dt^2-\frac 1{\left( 1-\frac{r_o%
}r\right) ^2}dr^2-r^2\left( d\theta ^2+\sin ^2\theta d\phi
^2\right).
\end{equation}
Klein-Gordon equation in a curved space time is,
\begin{equation}
\label{b}\left[ \frac 1{\sqrt{-g}}\partial _\kappa \left(
\sqrt{-g}g^{\kappa \nu }\partial _\nu \right) +\mu ^2\right] \Phi
\left( r,\theta,\phi,t\right) =0.
\end{equation}
Thus, the above equation becomes
\begin{equation}
\label{c}\left[ \frac 1{\left( 1-\frac{r_o}r\right) ^2}\frac{\partial ^2}{%
\partial t^2}-\frac 1{r^2}\frac \partial {\partial r}\left( r-r_0\right)
^2\frac \partial {\partial r}-\frac 1{r^2\sin \theta }\frac \partial
{\partial \theta }\sin \theta \frac \partial {\partial \theta
}-\frac 1{r^2\sin ^2\theta }\frac{\partial ^2}{\partial \phi ^2}+\mu
^2\right] \Phi \left( r,\theta,\phi,t\right) =0.
\end{equation}
Writing,
\begin{equation}
\label{d}\Phi \left( r,\theta,\phi,t\right) =\exp \left( -\imath
\epsilon t\right) Y_{lm}\left( \theta ,\phi \right) \Phi _l(r),
\end{equation}
where $\epsilon $, $l$, $m$ are energy, momentum and its projection,
while $ \Phi _l(r)$ is a radial function. Substituting Eq.(\ref{d})
in Eq.(\ref{c}) we get,
\begin{equation}
\label{e}\Phi _l^{\prime \prime }(r)+\frac 2{r-r_0}\Phi _l^{\prime
}(r)+\left( \frac{\epsilon ^2r^4}{\left( r-r_0\right) ^4}-\frac{\mu ^2r^2}{%
\left( r-r_0\right) ^2}-\frac{l\left( l+1\right) }{\left( r-r_0\right) ^2}%
\right) \Phi _l(r)=0.
\end{equation}
To study the scattering problem, we divide the space time outside
the black hole into 3 regions \cite{pp}. We start the three
different regions starting from the horizon as shown below.
\subsection{Region 1: $r\rightarrow r_{0}$}
Now we solve the wave equation in the vicinity of horizon, i.e., as $%
r\rightarrow r_0$, using the WKB approximation $%
\Phi =\exp \left({-\imath \int k(r)dr}\right)$, in Eq.(\ref{e}) and equating the real part we will get the radial wave number $%
k(r,l,\varepsilon )$ from the corresponding equation of motion:
therefore,
\begin{equation}
\label{c41}k\left( r\right) =\pm \left[ \epsilon ^2-\left( \frac{L^2}{r^2}%
-\mu ^2\right) \left( 1-\frac{r_0}r\right) ^2\right] ^{\frac
12}\frac{r^2}{ \left( r-r_0\right) ^2} .
\end{equation}
Thus near the horizon, i.e, as $r\rightarrow r_0$ ,
\begin{equation}
\label{o}k\left( r \rightarrow r_{0}\right) =\pm \frac{\epsilon r_0^2}{%
\left( r-r_0\right) ^2}.
\end{equation}
Therefore, the wave function in the region $r\rightarrow r_{b}$ can
be written as,
\begin{equation}
\label{lm1}\Phi_l(r)=\exp \left({\mp \imath \int \frac{%
\epsilon r_0^2dr}{\left( r-r_0\right) ^2}}\right)=\exp \left( \mp \imath \frac{\epsilon r_0^2}{%
r_0-r}\right),
\end{equation}

i.e.,
\begin{equation}
\label{p}\Phi_l\left( r\right) =\exp \left( \mp \imath \frac{\epsilon r_0^2}{%
r_0-r}\right).
\end{equation}
Let us describe the radial motion with the help of the wave function
$\Phi_l\left( r\right) $. Using Eq.(\ref{lm1}) the wave function in
the vicinity of horizon can be written, assuming that the wave gets
reflected at the horizon and having a reflection coefficient $|R|$,
as
\begin{equation}
\label{p1}\Phi_l\left( r\right) =\exp \left( -\imath \frac{\epsilon r_0^2}{%
r_0-r}\right) +\mid R\mid \exp \left( +\imath \frac{\epsilon r_0^2}{r_0-r}%
\right).
\end{equation}

\subsection{Region 2: $r>r_{0}$}
This region is considered to be sufficiently away from the horizon,
but not very far away from $r=r_{0}$. Assuming the terms in energy
and momentum in Eq.(\ref{e}) are very small compared to other terms,
and considering the s wave case, we will get the equation as,
\begin{equation}
\Phi_0^{\prime \prime }(r)+\frac 2{r-r_0}\Phi_0^{\prime }(r)=0.
\end{equation}
Therefore,
\begin{equation}
\Phi_0^{\prime }(r)=\frac C{\left( r-r_0\right) ^2},
\end{equation}
where $C$ is the constant of integration. Thus, the wave function
will be of the form,
\begin{equation}
\label{m}\Phi_0(r)=-\frac\alpha {\left( r-r_0\right) }+\beta.
\end{equation}
\subsubsection{Comparing Regions 1 and 2}
Here we compare the wave functions in regions 1 and 2. Eq.(\ref{p1})
can be written for s wave as,
\begin{equation}
\label{q}\Phi_0\left( r\right) =1-\imath \frac{\epsilon
r_0^2}{r_0-r}+\mid
R\mid \left( 1+\imath \frac{\epsilon r_0^2}{r_0-r}\right) =1+\mid R\mid +%
\frac{\left( 1-\mid R\mid \right) \imath \epsilon r_0^2}{r-r_0}.
\end{equation}
Comparing Eq.(\ref{q}) with the wave function in region 2 given by
Eq.(\ref{m}) we will get,
\begin{equation}
\label{r}\alpha =-\left( 1-\mid R\mid \right) \imath \epsilon r_0^2\text{, }%
\beta =1+\mid R\mid.
\end{equation}
\subsection{Region 3: $r>>r_{0}$}
This region is very far away from the horizon. Since, $r$ is very
large, we rewrite terms containing energy and momentum in
Eq.(\ref{e}) as,
\begin{equation}
\label{f}\frac{\epsilon ^2r^4}{\left( r-r_0\right) ^4}=\epsilon ^2+\frac{%
4\epsilon ^2r_0}{r-r_0}+\frac{4\epsilon ^2r_0^2}{\left( r-r_0\right) ^2}+%
\frac{\epsilon ^2r_0^2\left( r+r_0\right) }{\left( r-r_0\right) ^3}+\frac{%
\epsilon ^2r_0^2r^2}{\left( r-r_0\right) ^4},
\end{equation}
and
\begin{equation}
\label{g}\frac{\mu ^2r^2}{\left( r-r_0\right) ^2}=\mu ^2+\frac{2\mu ^2r_0}{%
\left( r-r_0\right) }+\frac{\mu ^2r_0^2}{\left( r-r_0\right) ^2}.
\end{equation}
Thus,
\begin{eqnarray}
\label{h1}\frac{\epsilon ^2r^4}{\left( r-r_0\right) ^4}-\frac{\mu ^2r^2}{%
\left( r-r_0\right) ^2}=\epsilon ^2-\mu ^2+\frac{\left( 2\epsilon
^2-\mu ^2\right) 2r_0}{r-r_0}+\frac{\left( 4\epsilon ^2-\mu
^2\right) r_0^2}{\left( r-r_0\right) ^2}+\frac{\epsilon
^2r_0^2\left( r+r_0\right) }{\left( r-r_0\right) ^3}+\frac{\epsilon
^2r_0^2r^2}{\left( r-r_0\right) ^4}, \nonumber \\ =p^2+\frac{\left(
p^2+\epsilon ^2\right) 2r_0}{r-r_0}+\frac{\left( p^2+3\epsilon
^2\right) r_0^2}{\left( r-r_0\right) ^2}+\frac{\epsilon
^2r_0^2\left( r+r_0\right) }{\left( r-r_0\right) ^3}+\frac{\epsilon
^2r_0^2r^2}{\left( r-r_0\right) ^4},
\end{eqnarray}
where $p^2=\epsilon ^2-\mu ^2$. Substituting Eq.(\ref{h1}) in
Eq.(\ref{e}) we get,
\begin{eqnarray}
\label{i}\Phi _l^{\prime \prime }(r)+\frac 2{r-r_0}\Phi _l^{\prime
}(r)+\left( p^2+\frac{\left( p^2+\epsilon ^2\right) 2r_0}{r-r_0}+\frac{%
\left( p^2+3\epsilon ^2\right) r_0^2}{\left( r-r_0\right)
^2}+\frac{\epsilon ^2r_0^2\left( r+r_0\right) }{\left( r-r_0\right)
^3}\nonumber\right .\\
 \left.+\frac{\epsilon ^2r_0^2r^2}{\left( r-r_0\right)
^4}-\frac{l\left( l+1\right) }{\left( r-r_0\right) ^2}\right) \Phi
_l(r)=0.
\end{eqnarray}
At $r\gg r_0$ we get,
\begin{equation}
\label{j}\Phi _l^{\prime \prime }(r)+\frac 2r\Phi _l^{\prime }(r)+\left( p^2+%
\frac{\left( p^2+\epsilon ^2\right) 2r_0}r-\frac{l\left( l+1\right) }{r^2}%
\right) \Phi _l(r)=0.
\end{equation}
Here the Coulomb charge is $Z=\left( p^2+\epsilon ^2\right) r_0$.
The solution to this equation will be,
\begin{equation}
\label{k}\Phi _l(r)=\frac 1r\left( A_l\exp \left( \imath z\right)
+B_l\exp \left( -\imath z\right) \right),
\end{equation}
where $z=pr-\frac{l\Pi }2+\nu \ln 2pr+\delta _t^{(c)}$, $\delta
_t^{(c)}=\arg \Gamma \left( l+1-\nu \right)$ and $\nu
=\frac Zp$. Here the wave equation can be considered as governed by Coulomb problem. Introducing regular $%
F\left( r\right) $ and singular $G\left( r\right)$, solution of
Coulomb problem, one can present the wave function as a linear
combination:
\begin{equation}
\label{s}\Phi_l(r)=\frac 1r\left( aF_l\left( r\right) +bG_l\left(
r\right) \right).
\end{equation}
In the asymptotic limit, $r\rightarrow \infty $, the Coulomb
functions takes the forms, $F_l\left( r\right) =\sin z$, $G_l\left(
r\right) =\cos z$ where $z=pr-\frac{l\Pi }2+\nu \ln 2pr+\delta
_t^{(c)}$, thus  Eq.(\ref{s}) will be in the form:
\begin{equation}
\label{es10}\Phi_l\left( r\right) =\frac 1r(a\sin z+b\cos z),
\end{equation}
But we know that for l=0 \cite{ee},
\begin{equation}
F_0\left( r\right) =cpr\,\text{, }G_0\left( r\right) =\frac 1c,
\end{equation}
where,
\begin{equation}
c^2=\frac{2\pi\nu}{1-2\pi\nu}.
\end{equation}
Thus Eq.(\ref{s}) for s wave becomes,
\begin{equation}
\label{s1}\Phi_0(r)=acp+\frac b{cr}.
\end{equation}

\subsubsection{Comparing Regions 2 and 3}
By neglecting higher powers of $\frac 1r$ Eq.(\ref{m}) can be
written as,
\begin{equation}
\label{t}\Phi_0(r)=-\frac \alpha {r\left( 1-\frac{r_0}r\right)
}+\beta \simeq- \frac \alpha r+\beta,
\end{equation}
and using Eq.(\ref{s1}), Eq.(\ref{t}) and Eq.(\ref{r}) we get,
\begin{equation}
\label{u}\text{ }a=\frac{1+\mid R\mid }{pc},b=\left( 1-\mid R\mid
\right) \imath \epsilon r_0^2c.
\end{equation}
\subsection{Absorption cross section of RN extremal black hole}
The two terms in Eq.(\ref{k}) represents the incoming and outgoing
waves. The $S$ matrix can be written as the ratio of coefficient of
the incoming and outgoing waves ($A_l$ and $B_l$), Therefore,
\begin{equation}
\label{l}S_l=\left( -1\right) ^{l+1}\frac{A_l}{B_l}\exp \left(
2\imath
\partial _l\right).
\end{equation}
We are here considering low energy absorption cross section for
$l=0$ (s wave). Thus from Eq.(\ref{k}) and Eq.(\ref{s}), we will
find coefficients $A_0$ and $B_0$ as,
\begin{equation}
\label{v}A_0=\frac{a+\imath b}{2\imath }\text{, }B_0=\frac{-a+\imath b}{%
2\imath }.
\end{equation}
Using Eq.(\ref{u}) we get,%
\begin{equation}
A_0=\frac{\left[ 1+\mid R\mid -\epsilon c^2p\left( 1-\mid R\mid
\right) r_o^2\right] }{2\imath pc},
\end{equation}
and%
\begin{equation}
B_0=-\frac{\left[ 1+\mid R\mid +\epsilon c^2p\left( 1-\mid R\mid
\right) r_0^2\right] }{2\imath pc}.
\end{equation}
Thus the S-matrix for the s-wave is given by,
\begin{eqnarray}
S_0=-\frac{A_0}{B_0}\exp \left( 2\imath \partial _0\right)
=\frac{1+\mid R\mid -\epsilon c^2p\left( 1-\mid R\mid \right)
r_o^2}{1+\mid R\mid -\epsilon c^2p\left( 1-\mid R\mid \right)
r_o^2}\exp \left( 2\imath \partial _0\right)\nonumber
\\=\frac{1-\epsilon c^2pr_o^2\eta }{1+\epsilon c^2pr_o^2\eta }\exp
\left( 2\imath
\partial _0\right),
\end{eqnarray}
where $\eta =\frac{1+\mid R\mid }{1-\mid R\mid }$. The absorption
cross section in the low energy limit is given by \cite{ee},
\begin{equation}
\label{x}\sigma _{abs}=\frac \pi {p^2}\left( 1-\mid S_o\mid
^2\right) =\frac \pi {p^2}\frac{4c^2\epsilon pr_0^2\eta }{\left(
1+\epsilon c^2pr_0^2\eta \right) ^2},
\end{equation}
taking $p=\epsilon v$, we get Eq.(\ref{x}) as,
\begin{equation}
\label{y}\sigma _{abs}=\frac{4\pi c^2r_0^2\eta }{v\left( 1+\epsilon
^2c^2vr_0^2\eta \right) ^2}.
\end{equation}

\section{ Schwarzschild-de Sitter black hole - Extremal case}
The metric of a SdS space time is given by,
\begin{equation}
\label{es1}ds^2=f(r)dt^2- f(r)^{-1}dr^2-r^2\left( d\theta ^2+\sin
^2\theta d\phi ^2\right),
\end{equation}
where $f(r) =1-\frac{2m}{r}-\frac{\Lambda}{3}r^{2}$ with $\Lambda>0$
and $m>0$. For $0<9\Lambda m^{2}<1$ there exist two positive roots
$r_{+}$ and $r_{++}$ of $f(r)$ such that $0<2m<r_{+}<3m<r_{++}$. The
roots $r_{+}=\frac{2}{\sqrt{\Lambda}}\cos \left( \frac{\alpha
}{3}+\frac{4\pi }{3}\right)$ with $\cos \alpha =-3m\sqrt{\Lambda}$,
describes the black hole event horizon, and the root
$r_{++}=\frac{2}{\sqrt{\Lambda}}\cos \left( \frac{\alpha
}{3}\right)$ localizes the cosmological event horizon. As $\Lambda$
approaches its extremal value, i.e,
$\Lambda\rightarrow\frac{1}{9m^{2}}$, the position of the black hole
horizon $r_{+}$ monotonically increases and the cosmological horizon
$r_{++}$ decreases to a common value at $r=3m$. Here we analyze this
extreme case of the SdS black hole which is characterized by the
condition $9\Lambda m^{2}=1$. In this case the $f(r)$ becomes
\cite{jpod99},
\begin{equation}
\label{es2}f(r) =-\frac 1{27m^2r}(r-3m)^2(r+6m).
\end{equation}
Applying this in  Klein Gordon equation, we will get,
\begin{eqnarray}
\label{es3}\left[ \frac{-27m^2r}{(r-3m)^2(r+6m)}\frac{\partial
^2}{\partial t^2}+\frac 1{27m^2r^2}\frac \partial {\partial
r}r(r-3m)^2(r+6m)\frac
\partial {\partial r}\nonumber\right .\\
 \left.-\frac 1{r^2\sin \theta }\frac \partial {\partial
\theta }\sin \theta \frac \partial {\partial \theta }-\frac
1{r^2\sin ^2\theta }\frac{\partial ^2}{\partial \phi ^2}+\mu
^2\right] \psi \left( r,\theta ,\phi ,t\right) =0.
\end{eqnarray}
Real part is separated out by,
\begin{equation}
\label{es4}\psi \left( r,\theta ,\phi ,t\right) =\exp \left( -\imath
\epsilon t\right) Y_{lm}\left( \theta ,\phi \right) \Phi _l(r).
\end{equation}
where $\epsilon $, $l$, $m$ are energy, momentum and its projection, while $%
\Phi _l(r)$ is a radial function. Substituting Eq.(\ref{es4}) in
Eq.(\ref {es3}) we get,
\begin{eqnarray}
\label{es5}\Phi _l^{\prime \prime }(r)+(\frac 1r+\frac 2{r-3m}+\frac
1{r+6m})\Phi _l^{\prime }(r)+\left( \frac{\epsilon
^227^2m^4r^2}{\left(
r-3m\right) ^4(r+6m)^2}\nonumber\right .\\
\left.+\frac{\mu ^227m^2r}{\left( r-3m\right) ^2(r+6m)}-%
\frac{l\left( l+1\right) 27m^2r}{\left( r-3m\right) ^2(r+6m)}\right)
\Phi _l(r)=0.
\end{eqnarray}
Let us describe the scattering of scalar waves by SdS extremal black
hole with the help of Eq.(\ref{es5}). We will find solution of the
wave equation in different regions outside the horizon.
\subsection{Region 1: $r\rightarrow 3m$}
This is the region very near to the horizon and is the last limit
for a wave to reflect from the horizon. i.e, the wave function
$\Phi(r)$ in this region (i.e., as $ r\rightarrow 3m$) will contain
incident and reflected waves. To find $\Phi(r)$ we use WKB
approximation and write $\Phi=\exp\left({-\imath\int
k(r)dr}\right)$, which will lead to,
\begin{equation}
\label{es17}k\left( r \rightarrow 3m\right) =\pm \frac{\epsilon
27m^2\times 3m}{ (r-3m)^2(3m+6m)}=\pm \frac{9m^2\epsilon }{\left(
r-3m\right) ^2}=\pm \frac \xi {\left( r-3m\right) ^2}.
\end{equation}
Therefore the wave function in the region $r\rightarrow 3m$ can be
written as,
\begin{equation}
\label{lm}\Phi_l(r)=\exp \left({\mp \imath \int \frac{\xi dr}{\left(
r-3m\right) ^2}}\right)=\exp \left(\pm \imath \frac{\xi}{\left(
r-3m\right) ^2}\right),
\end{equation}
where $\xi =\epsilon 9m^2$. Therefore, the wave function in the
vicinity of horizon can be written, assuming that the wave gets
reflected at the horizon, as
\begin{equation}
\label{es19}\Phi_l\left( r\right) =\exp \left( -i\frac \xi
{3m-r}\right) +\mid R\mid \exp \left( +i\frac \xi {3m-r}\right).
\end{equation}
where $|R|$ represents the reflection coefficient.
\subsection{Region 2: $r>3m$}
Now we consider the second region where $r>3m$. As in section(2.3),
here also we neglect the energy and momentum terms in Eq.(\ref{es5})
and for the s wave the resulting equation will be,
\begin{equation}
\label{es13}\Phi_0^{\prime \prime }(r)+(\frac 1r+\frac 2{r-3m}+\frac
1{r+6m})\Phi_0^{\prime }(r)=0,
\end{equation}
i.e,
\begin{equation}
\label{es14}\ln \Phi_0^{\prime }(r)=-2\ln (r-3m)-\ln r-\ln (r+6m).
\end{equation}
Therefore
\begin{equation}
\label{es15}\Phi_0(r)=-\frac \alpha {r-3m}+\beta.
\end{equation}

\subsubsection{Comparing Regions 1 and 2}
As in Section(2.3.1) here also we can compare the solution in the
regions 1 and 2. Then Eq.(\ref{es19}) for s wave, can be written as,
\begin{equation}
\label{es20}\Phi_0\left( r\right) =1-i\frac \xi {3m-r}+\mid R\mid
\left( 1+i\frac \xi {3m-r}\right) =1+\mid R\mid +\frac{i\xi (1-\mid
R\mid )}{r-3m}.
\end{equation}
Comparing  Eq.(\ref{es20})with Eq.(\ref{es15}) we get,
\begin{equation}
\label{es21}\alpha =-\left( 1-\mid R\mid \right) i\xi \text{, }\beta
=1+\mid R\mid.
\end{equation}

\subsection{Region 3: $r>>3m$}
In this region, the terms containing energy and momentum in
Eq.(\ref{es5}) can be simplified as,
\begin{eqnarray}
\label{es6}\frac{\epsilon ^227^2m^4r^2}{\left( r-3m\right) ^4(r+6m)^2}+\frac{%
\mu ^227m^2r}{\left( r-3m\right) ^2(r+6m)}\simeq \epsilon ^2-\mu ^2+\frac{%
(2\epsilon ^2-\mu ^2)27\times 2m^3}{\left( r-3m\right) ^2(r+6m)}
\nonumber \\
=p^2+\frac{(p^2+\epsilon ^2)27\times 2m^3}{\left( r-3m\right)
^2(r+6m)},
\end{eqnarray}
where $p$ is the momentum and is given by $p^2=\epsilon ^2-\mu ^2$.
Thus Eq.(\ref{es5})becomes,
\begin{equation}
\label{es7}\Phi _l^{\prime \prime }(r)+(\frac 1r+\frac 2{r-3m}+\frac
1{r+6m})\Phi _l^{\prime }(r)+\left( p^2+\frac{(p^2+\epsilon ^2)27\times 2m^3%
}{\left( r-3m\right) ^2(r+6m)^{}}-\frac{l\left( l+1\right)
27m^2r}{\left( r-3m\right) ^2(r+6m)}\right) \Phi _l(r)=0.
\end{equation}
Since we are considering s wave we take $l=0$. Now at region $r>>3m$
we can neglect higher powers of $\frac{1}{r}$ and we get the
equation as,
\begin{equation}
\label{es8}\Phi_0^{\prime \prime }(r)+(\frac 1r+\frac 2{r-3m}+\frac
1{r+6m})\Phi_0^{\prime }(r)+p^2\Phi_0(r)=0.
\end{equation}
Since $r+6m>r>r-3m$ we can approximate the coefficient of $\Phi_0
^{\prime }(r)$ as $\frac 2r$. Therefore at large distance we can
write the equation as,
\begin{equation}
\label{es9}\Phi_0^{\prime \prime }(r)+\frac 2r\Phi_0^{\prime
}(r)+p^2\Phi_0(r)=0.
\end{equation}
Solution of Eq.(\ref{es9}) is obtained as a combination of $sinz$
and $cosz$ using Frobenius method. Thus solution can be written as,
\begin{equation}
\label{es11}\Phi_0\left( r\right) =\frac 1r(A_0\exp iz+B_0\exp -iz),
\end{equation}
where $z=pr$. Here also, as in the previous case (section(2.3)), the
wave function can be written as a combination of two functions
$F\left( r\right) $ and $G\left( r\right) $. Therefore
\begin{equation}
\label{es22}\Phi_0(r)=\frac 1r\left( aF\left( r\right) +bG\left(
r\right) \right),
\end{equation}
but by comparing with the solutions obtained using Frobenious
method, we know that $F\left( r\right)=sinpr $ and $G\left(
r\right)=cospr $. From this we will deduce that,
\begin{equation}
\label{es12}A_0=\frac{a+ib}{2i},B_0=\frac{-a+ib}{2i}.
\end{equation}
On the other hand, since $\epsilon$ is low we can assume $pr<<1$ and
thus,
\begin{equation}
F\left( r\right) \simeq pr\,\text{, }G\left( r\right) \simeq1.
\end{equation}
Thus Eq.(\ref{es22}) will become,
\begin{equation}
\label{es23}\Phi_0(r)=ap+\frac br.
\end{equation}
\subsubsection{Comparing Regions 2 and 3}
By neglecting higher powers of $\frac 1r$ Eq.(\ref{es15}) can also
be written as,
\begin{equation}
\label{es24}\Phi_0(r)=-\frac \alpha {r\left( 1-\frac{3m}r\right)
}+\beta \simeq -\frac \alpha r+\beta,
\end{equation}
Thus Eq.(\ref{es24}) has the same form as Eq.(\ref{es23}). Therefore
comparing the coefficients we will get,
\begin{equation}
\label{es25}\text{ }a=\frac{1+\mid R\mid }p\text{, }b=i\left( 1-\mid
R\mid \right) \xi.
\end{equation}
\subsection{Absorption cross section of SdS extremal black hole}
Now we will find an expression for absorption cross section of SdS
extremal black hole. From Eq.(\ref{es12}) and Eq.(\ref{es25}) we
will get,
\begin{equation}
A_0=\frac{\left[ 1+\mid R\mid -\left( 1-\mid R\mid \right) \xi p\right] }{%
2\imath pc},
\end{equation}
and%
\begin{equation}
B_0=-\frac{\left[ 1+\mid R\mid +\left( 1-\mid R\mid \right) \xi p\right] }{%
2\imath pc}.
\end{equation}
S-matrix for the s-wave is given by,
\begin{equation}
S_0=-\frac{A_0}{B_0}\exp \left( 2\imath \partial _0\right)
=\frac{\left[ 1+\mid R\mid -\left( 1-\mid R\mid \right) \xi p\right]
}{\left[ 1+\mid R\mid +\left( 1-\mid R\mid \right) \xi p\right]
}\exp \left( 2\imath \partial _0\right).
\end{equation}
It can also be written as,
\begin{equation}
\label{es26}S_0=\frac{1-\xi p\eta }{1+\xi p\eta }\exp \left( 2\imath
\partial _0\right),
\end{equation}
where $ \eta =\frac{1-\mid R\mid }{1+\mid R\mid }$. The absorption
cross section in the low energy limit is given by,
\begin{equation}
\label{es27}\sigma _{abs}=\frac \pi {p^2}\left( 1-\mid S_o\mid
^2\right) =\frac \pi p\frac{4\xi \eta }{\left( 1+\xi p\eta \right)
^2},
\end{equation}
i.e,
\begin{equation}
\label{y}\sigma _{abs}=\frac{36\pi m^2\eta }{v\left( 1+9m^2\epsilon
^2c^2v\eta \right) ^2}.
\end{equation}

\section{Conclusion}
We  found the wave function $\Phi_{l} (r)$ in the vicinity of outer
horizon of RN extremal black hole i.e $r\rightarrow r_0$ for scalar
field using WKB approximation. We have also studied the behavior of
scattered scalar waves in the regions $r>r_0$ and $r\gg r_0$ in low
energy limit. By comparing the solutions in the 3 regions viz.,
$r\rightarrow r_0$, $r>r_0$ and $r\gg r_0$, we found the S-matrix
and the absorption cross section for RN extremal black hole by
s-wave in the lower energy limit. Similarly we found the wave
functions $\Phi_l (r)$ in the vicinity of outer horizon
$r\rightarrow 3m$ and in the regions $r>3m$ and $r\gg 3m$ of SdS
extremal black hole for scalar field. We have also studied the
behavior of scattered scalar waves in all these regions in low
energy limit. By comparing the solutions in the 3 regions viz.,
$r\rightarrow 3m$, $r>3m$ and $r\gg 3m$, we found the absorption
cross section for SdS extremal black hole by s-wave in the lower
energy limit.

\section*{Acknowledgments}
SR and VCK are thankful to U.G.C, New Delhi for financial support
through a Major Research Project. VCK wishes to acknowledge
Associateship of IUCAA, Pune. India.


\begin{thebibliography}{0}
\bibitem{ss} J. A. H. Futterman, F. A. Handler and R. A. Matzner, {\it Scattering from Black holes} (Cambridge University Press, Cambridge, 1988)
\bibitem{ii}  R. Penrose, {\it Riv. Nuovo Cimento} {\bf 1}, 252
(1969).
\bibitem{abaa} S. W. Hawking, {\it Commun. Math. Phys.} \textbf{43}, 199(1975).
\bibitem{pp}  W. G. Unruh, {\it Phys. Rev. D} {\bf 14}, 3251 (1976).
\bibitem{abab} N. Sanchez, {\it Phys. Rev. D} {\bf 16}, 937 (1977).
\bibitem{abac} N. Sanchez, {\it Phys. Rev. D} {\bf 18}, 1030 (1978).
\bibitem{abad} N. Sanchez, {\it Phys. Rev. D} {\bf 18}, 1798 (1978).
\bibitem{viv} G.'t Hooft, {\it gr-qc}/0406017
\bibitem{22} Mu-Lin Yan and Hua Bai, {\it gr-qc}/0401027
\bibitem{bw98} B. Wang, R. K. Su, P. K. N. Yu and E. C. M. Young, {\it Phys. Rev. D} {\bf 57}, 5284 (1998).
\bibitem{sbn} I. D. Novikov and V. P. Frolov, {\it Physics of Black holes} (Kluwer Academic publishers, 1989).
\bibitem{bb}  M. Yu. Kuchiev, {\it Phys. Rev. D} {\bf 69}, 124031 (2004)
\bibitem{sr}  R. Sini and V. C. Kuriakose, {\it Int. J. Mod. Phys D} {\bf 16}, 105 (2007).
\bibitem{sr2} R. Sini and V. C. Kuriakose, {\it gr-qc}/0708.3146 (Int. J. Mod. Phys D(to
appear)).
\bibitem{sr3} R. Sini and V.C. Kuriakose, {\it(Mod. Phys. Lett. A} (to appear)).
\bibitem{dd}  Feng-Li Lin, {\it hep-th}/9807084.
\bibitem{ee}  L. D. Landau and E. M. Lifshits, {\it Quantum Mechanics:
Non-Relativistic Theory} (Pergamon, New York, 1977).
\bibitem{jpod99} J. Podolsky, {\it Gen. Rel. Grav.} {\bf 31}, 1703 (1979).
\end{thebibliography}
\end{document}